\documentclass[a4paper,aps,11pt,nofootinbib,showpacs]{revtex4}
\usepackage{amsmath}
\usepackage{amssymb}
\usepackage{amsfonts}
\usepackage[dvips]{graphicx}

\begin{document}

%%%%%%%%%%%%%%%%%%%%%%%%%%%%%%%%%%%%%%%%
\title{\Large Chaos in Geodesic Motion around a Black Ring}
%{\today}

\hfill{OCU-PHYS 343}

\hfill{AP-GR 87}

%pacs-> Higher-dimensional black holes, black strings, and related objects 

\author{Takahisa Igata} 
\email{igata@sci.osaka-cu.ac.jp}
\author{Hideki Ishihara}
\email{ishihara@sci.osaka-cu.ac.jp}
\author{Yohsuke Takamori}
\email{takamori@sci.osaka-cu.ac.jp}
\affiliation{%
 Department of Mathematics and Physics,
 Graduate School of Science, Osaka City University,
 Osaka 558-8585, Japan}
%%%%%%%%%%%%%%%%%%%%%%%%%%%%%%%%%%%%%%%%

%%%%%%%%%%%%%%%%%%%%%%%%%%%%%%%%%%%%%%%%
\begin{abstract}
We study bound orbits of a free particle around a singly rotating black ring. 
We find there exists chaotic motion of a particle which is gravitationally 
bound to the black ring by using the Poincar\'e map. 
\end{abstract}
%%%%%%%%%%%%%%%%%%%%%%%%%%%%%%%%%%%%%%%%
\maketitle

\section
%\noindent
{Introduction}

Chaos is one of the characteristic behavior of non-linear dynamical systems. 
In the context of general relativity, there are two main issues concerning chaos. 
One is chaotic oscillations which 
generally occur in the early stage of the universe near the initial 
singularity\cite{BKL, cosmology}. 
The other is chaotic motion of particles around black holes. 
There appears chaotic behavior of charged particles 
around a magnetized black hole\cite{magnetized_BH}, 
particles around a gravitationally perturbed black hole\cite{perturbed_BH}, 
a spinning particle around a black hole in vacuum\cite{Suzuki:1997}, and particles 
around multi-black holes\cite{multi_BH}. 

Recently, general relativity in higher dimensions gathers much attention 
in relation to modern unified theories of interactions. 
Properties of the gravitational field depend on 
the spacetime dimensions critically. 
As for the cosmological models, the chaotic oscillations of the early universe 
disappear in higher dimensions\cite{KKcosmology}. 
As for the black holes, in five dimensions, exact solutions of a black ring 
with the horizon topology of 
S$^2\times$ S$^1$ are discovered by Emparan and Reall\cite{Emparan:2001wn} 
in addition to rotating black holes with the spherical horizon topology obtained by 
Myers and Perry\cite{Myers:1986un}.  

The geodesic motion of a test particle is one of the most important probes 
for spacetime geometry 
because it reveals the geometrical difference of the black ring and the black hole. 
It is known that Myers-Perry black holes in any dimensions allow separation 
of variables in the Hamilton-Jacobi equation for 
geodesics\cite{Frolov_etal} 
as same as the Kerr black hole in four dimensions. 
This occurs because of the existence of a rank-2 Killing tensor in addition 
to Killing vectors generating isometries. 
However, the separation of variables in the black ring geometry does not occur 
by the ring coordinates\cite{Hoskisson:2007zk, Durkee:2008an}. 
As another interesting difference, the black rings have stable bound orbits of 
a particle\cite{Igata:2010ye}, while the black holes in five dimensions do not. 
This comes from the difference of shapes of black objects. 

If the particle motion bounded in a finite region are not integrable, 
the following natural question arises. Is the particle motion chaotic? 
The non-separability of variables in the black ring geometry suggests that 
there is no additional constant of motion 
except constants associated with the Killing vectors. 
However, we cannot conclude the absence of additional constant of motion 
immediately 
because the Hamilton-Jacobi method depends on the choice of variables. 
In this report, we show the black ring geometry has chaotic bound orbits 
by using the Poincar\'e map. Appearance of the chaos implies the absence of 
additional constant of motion in the black ring metric.

%%%%%%%%%%%%%%%%%%%%%%%%%%%%%%%%%%%%%%%%%%%%%%%%%%%%%%%%%%%%%%%%%%%%%
\section
%\noindent
{ Geometry of the Black Ring }
%%%%%%%%%%%%%%%%%%%%%%%%%%%%%%%%%%%%%%%%%%%%%%%%%%%%%%%%%%%%%%%%%%%%%

In terms of ring coordinates $(t,x,y,\phi,\psi)$, 
the black ring metric is given by  
\begin{align}
	ds^2 =& - \frac{F(y)}{F(x)}
		\left(dt-CR\frac{1+y}{F(y)} d\psi\right)^2 
\cr&
	+\frac{R^2}{(x-y)^2}F(x)
		\left(- \frac{G(y)}{F(y)}d\psi^2- \frac{dy^2}{G(y)}
		+ \frac{dx^2}{G(x)}+ \frac{G(x)}{F(x)}d\phi^2
      \right),
\label{eq:BlackRing}
\end{align}
where
\begin{align}
	&F(\xi)=1+\lambda\xi,
\quad
	G(\xi)=(1-\xi^2)(1 + \nu \xi),
\\
	&C=\sqrt{\lambda(\lambda-\nu)\frac{1+\lambda}{1-\lambda}}, 
\end{align}
where the parameter $R$ denotes the radius of the black ring, 
and  $\lambda$  
and $\nu$ characterize the rotation velocity and the thickness of the ring, 
respectively.
The ranges of the parameters are
\begin{align}
	0<R, \quad 0<\nu\leq\lambda<1, 
\label{eq:nu}
\end{align}
and the ranges of the ring coordinates are given by 
\begin{align}
	-\infty \leq y \leq -1, \quad
	-1 \leq x \leq 1. 
\end{align} 

In the black ring metric \eqref{eq:BlackRing}, $y = - 1/\nu$ 
is the position of the event horizon which has the topology of 
S$^2 \times$ S$^1$. 
%The singularity at $y= - \infty$ corresponds to the curvature singularity. 
The metric admits three Killing vectors, $\partial_t, \partial_\psi$, and 
$\partial_\phi$. The ring axis, fixed points of the rotation generated 
by $\partial_\psi$, is $y=-1$, and the equatorial plane, fixed points of 
the rotation generated by $\partial_\phi$, is $x=\pm1$. 
The ergosurface exists at $y = -1/\lambda$, i.e., 
the Killing vector $\partial_t$, which is timelike 
at the spatial infinity, becomes null there. 
In terms of regularity condition at the ring axis and the equatorial plane, 
$\lambda$ has to be chosen as
\begin{align}
	\lambda=\frac{2\nu}{1+\nu^2}, 
\label{eq:regularity}
\end{align}
then the regular black ring solutions have two free parameters $R$ and $\nu$.

%%%%%%%%%%%%%%%%%%%%%%%%%%%%%%%%%%%%%%%%%%%%%%%%%
\section
%\noindent
{ Particle Motion around the Black Ring}
%%%%%%%%%%%%%%%%%%%%%%%%%%%%%%%%%%%%%%%%%%%%%%%%%

The Hamiltonian of a free particle with mass $m$ is generally given by
\begin{align}
	H = \frac{N}{2}\left(g^{\mu\nu} p_{\mu}p_{\nu} + m^2\right), 
\end{align}
where $N$ is the Lagrange multiplier and $p_{\mu}$ is the canonical momentum. 
In the case of particle motion around the black ring metric \eqref{eq:BlackRing}, 
since $t, \psi$, and $\phi$ are cyclic coordinates, then the conjugate momenta 
$p_t, p_\psi$, and $p_\phi$ are constants of motion. 
Then, the geodesic Hamiltonian is reduced in the form
\begin{align}
	H = \frac{N}{2}\left[
		g^{xx}p_x^2 +g^{yy}p_y^2
 		+ E^2 \left(U_{{\rm eff}} + \frac{m^2}{E^2}\right)\right], 
\label{eq:Hamiltonian}
\end{align}
where 
\begin{align}
	U_{{\rm eff}}
	 =& g^{tt} + g^{\phi\phi}l_{\phi}^2
		 + g^{\psi\psi}l_{\psi}^2 - 2g^{t\psi}l_{\psi}
\end{align}
with
\begin{align}
	&g^{tt} = - \frac{F(x)}{F(y)} - \frac{C^2(x-y)^2(y+1)^2}{G(y)F(x)F(y)}, \quad
	g^{xx} = \frac{(x-y)^2}{R^2}\frac{G(x)}{F(x)},\quad
	g^{yy} = - \frac{(x-y)^2}{R^2} \frac{G(y)}{F(x)},
\cr
	&g^{\phi\phi} = \frac{(x-y)^2}{R^2 G(x)},\quad 
	g^{\psi\psi} = - \frac{F(y)(x-y)^2}{R^2 G(y) F(x)},\quad
	g^{t\psi} = - \frac{C(x-y)^2(y+1)}{R G(y) F(x)}, 
\label{eq:guu}
\end{align}
and $E=- p_t$, $l_{\phi}=p_{\phi}/E$, and $l _{\psi}=p_{\psi}/E$ are constants. 

By variation of the geodesic action with $N$, 
we obtain the Hamiltonian constraint condition
\begin{align}
	g^{xx}p_x^2 +g^{yy}p_y^2
 		+ E^2 \left(U_{{\rm eff}} + \frac{m^2}{E^2}\right)=0. 
\label{eq:Hamiltonian_Constraint}
\end{align}

In what follows, to give more intuitive pictures of particle motion, 
we use $\zeta$-$\rho$ coordinates which are defined as
\begin{align}
	\zeta = R \frac{\sqrt{y^2-1}}{x-y}, \quad
	\rho = R \frac{1-x^2}{x-y}. 
\end{align}
In this coordinates, the flat metric takes the form
\begin{equation}
	ds^2 = -dt^2 + d\zeta^2 +\zeta^2 d\psi^2 +d\rho^2 +\rho^2 d\phi^2. 
\end{equation}
The ring axis and the equatorial plane 
correspond to $\zeta=0$ and $\rho = 0$, respectively, 
and the horizon of the black ring $y=-1/\nu$ is represented by a circle on the 
equatorial plane. 
The effective potential $U_{\rm eff}$ is a function of $\zeta$ and $\rho$ with 
the parameters $\nu$, $l_\psi$, and $l_\phi$.  

As is shown in the previous work\cite{Igata:2010ye}, 
if $l_{\phi}$ and $l_{\psi}$ are chosen in a suitable range, 
the effective potential $U_{\rm eff}$ has a local minimum at a point, 
say $(\zeta_{\rm s}, \rho_{\rm s})$, 
i.e., there exist stable bound orbits around the black ring. 
The projection of each orbit on a time slice is 
a toroidal spiral curve on the two-dimensional torus, 
direct product of S$^1$ with radius $\zeta_{\rm s}$ and S$^1$ with radius 
$\rho_{\rm s}$.
In the case that $l_{\psi}=0$, a potential minimum appears on the ring 
axis $\zeta=0$. 
A minimum point on the ring axis $(\zeta_{\rm s}=0,\rho_{\rm s})$ implies a 
stable circular orbit of the radius $\rho_{\rm s}$ on the ring axis. 
There also exist potential minima off the ring axis for some $l_\phi$. 
It means that the orbits with $l_{\psi}=0$ can take toroidal spiral shapes 
because of dragging by the rotation of black rings. 

%%%%%%%%%%%%%%%%%%%%%%%%%%%%%%%%%%%%%%%%%%%%%%%%%%%%%%%%
\section
%\noindent
{Chaotic Motion}
%%%%%%%%%%%%%%%%%%%%%%%%%%%%%%%%%%%%%%%%%%%%%%%%%%%%%%%%

\begin{figure}[htbp]
\begin{tabular}{ccc}
\begin{minipage}{0.3\hsize}
\begin{center}
{(a) $E=0.941$}
 \includegraphics[width=5cm,clip]{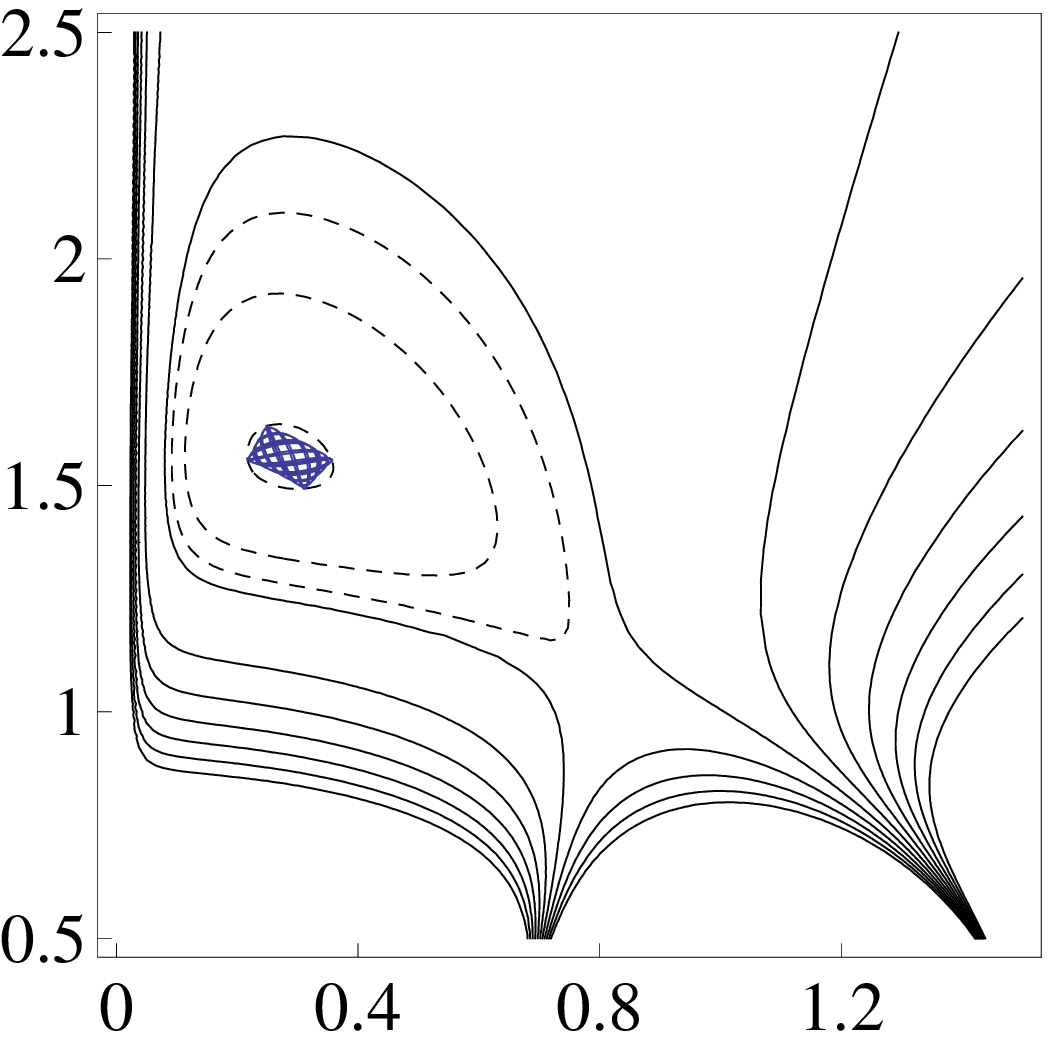}
\end{center}
\end{minipage}
\hspace{5mm}
\begin{minipage}{0.3\hsize}
\begin{center}
{(b) $E=0.947$}
 \includegraphics[width=5cm,clip]{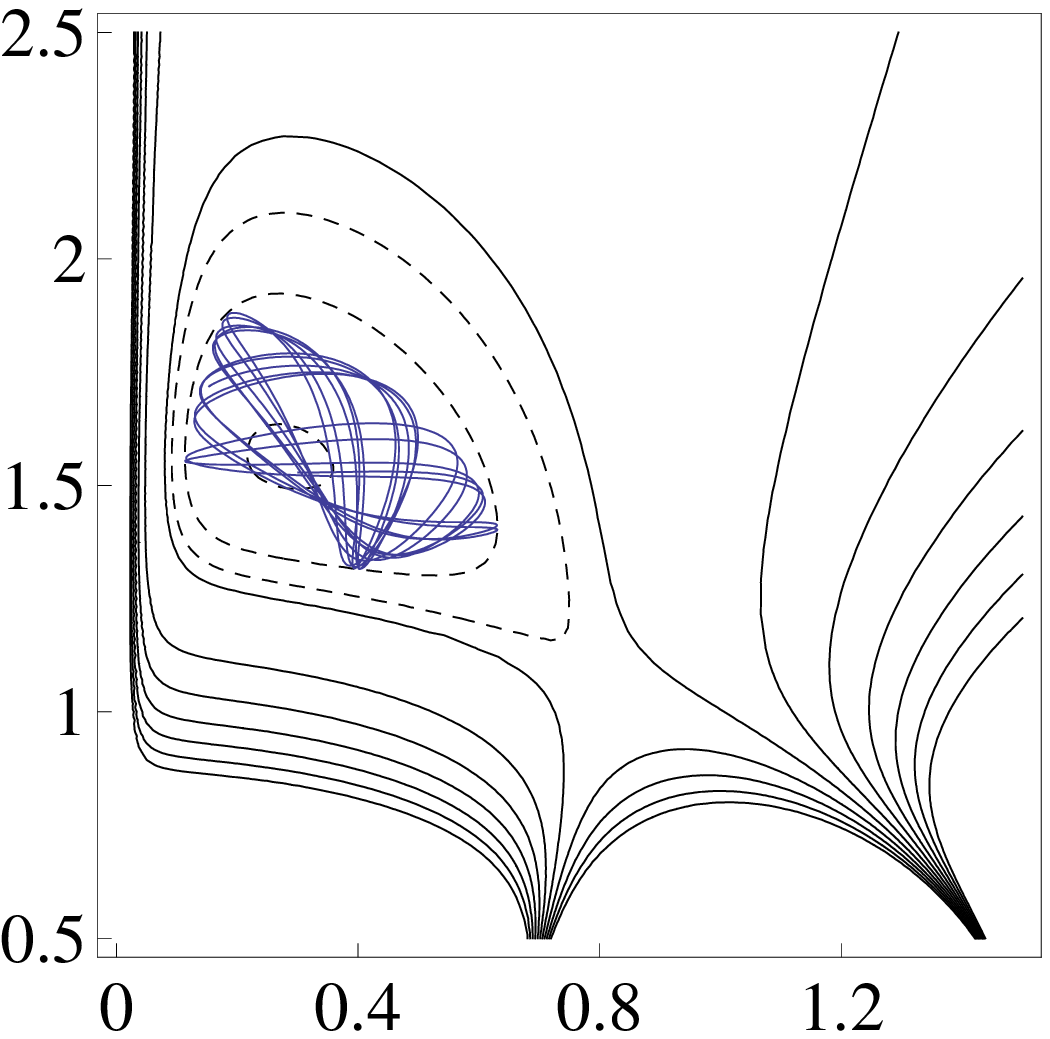}
 \end{center}
\end{minipage}
\hspace{5mm}
\begin{minipage}{0.3\hsize}
 \begin{center}
{(c) $E=0.952$}
 \includegraphics[width=5cm,clip]{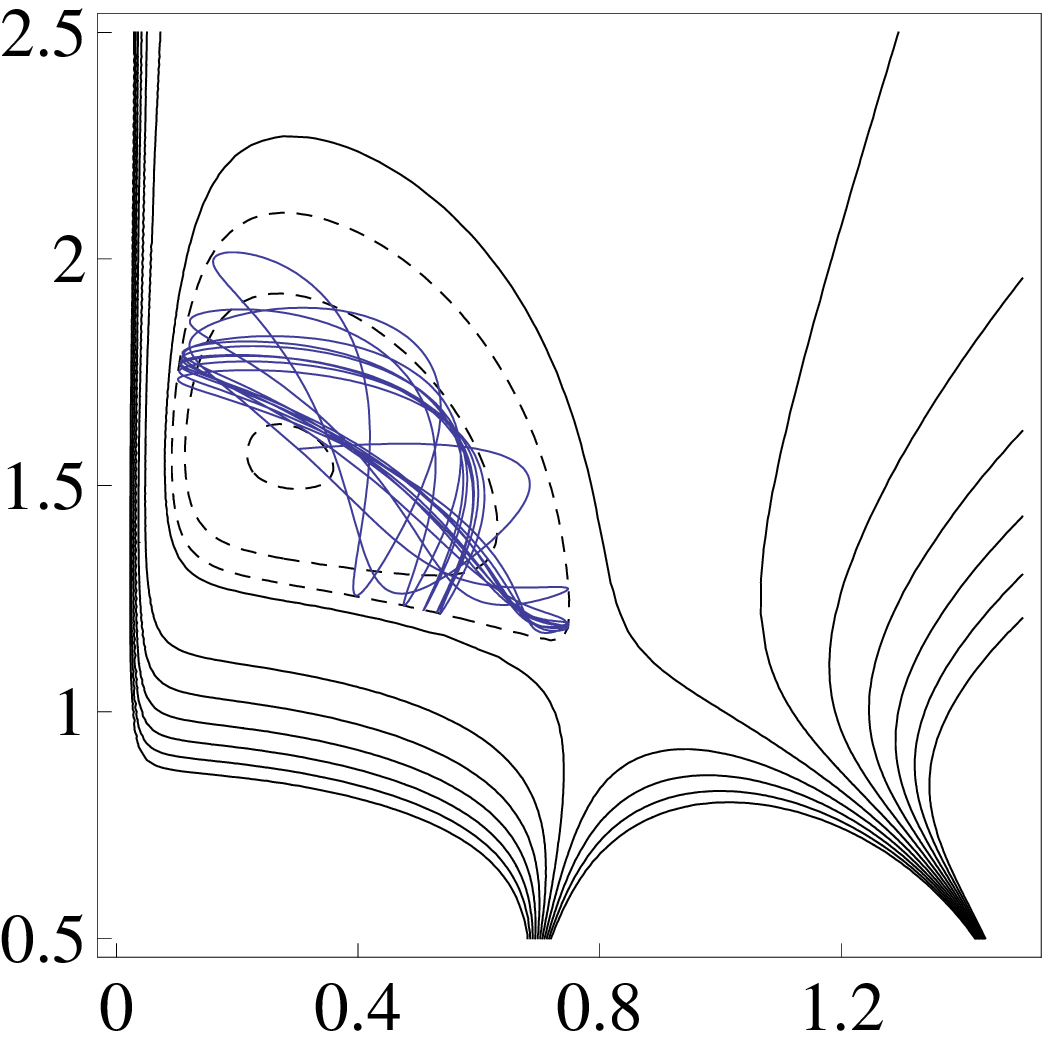}
 \end{center}
\end{minipage}
\end{tabular}
\caption{
 \label{dynamical_orbits}
Orbits of a particle are plotted on the $\zeta$-$\rho$ plane with contours 
of the effective potential $U_{\rm eff}$. 
The horizontal axis denotes $\zeta$ and the vertical axis denotes $\rho$. 
The parameters are set as $\nu=0.4$ and $R=1$ for the black ring geometry, and  
$l_{\phi} = 1.52$ and $l_{\psi} = 0.02$ for constants of motion. 
Energies are (a) $E=0.941$, (b) $E=0.947$, and (c) $E=0.952$. 
The energy levels of the orbits are shown by broken closed curves. 
}
\end{figure}

Now, we observe appearance of chaotic behavior of bound orbits around 
the black ring. 
We consider dynamical geodesic motion bounded in a finite region. 
Such orbits exist near the stable bound orbits. 
In Fig.\ref{dynamical_orbits}, we show typical orbits in the $\zeta$-$\rho$ plane 
with contours of $U_{\rm eff}$ by solving the equations of motion numerically. 
We find a saddle point of $U_{\rm eff}$ between the local minimum and the horizon 
(see Fig.\ref{dynamical_orbits}). The particle motion with the energy $E$ 
in the range 
\begin{align}
	E_{\rm s} \leq E < E_{\rm u}
\end{align} 
is bounded in a finite region around the local minimum, 
where $E_{\rm s}$ and $E_{\rm u}$ are energy levels of 
the local minimum (stable point) at $(\zeta_{\rm s}, \rho_{\rm s})$ and the 
saddle point (unstable point) of $U_{\rm eff}$, respectively. 
If the energy of the particle is  a little bit larger than $E_{\rm s}$ 
such that the particle orbit is confined in a vicinity of 
the local minimum, the orbit makes a Lissajous figure. 
As the energy $E$ increases, the Lissajous figure is deformed, and in the case that 
the energy becomes as large as $E_{\rm u}$ 
such that particle can approach to the saddle point of $U_{\rm eff}$, 
the orbits become complicated and irregular. 
\begin{figure}[h]
\begin{tabular}{ccc}
\begin{minipage}{0.3\hsize}
 \begin{center}
{(a) $E=0.941$}
 \includegraphics[width=5cm,clip]{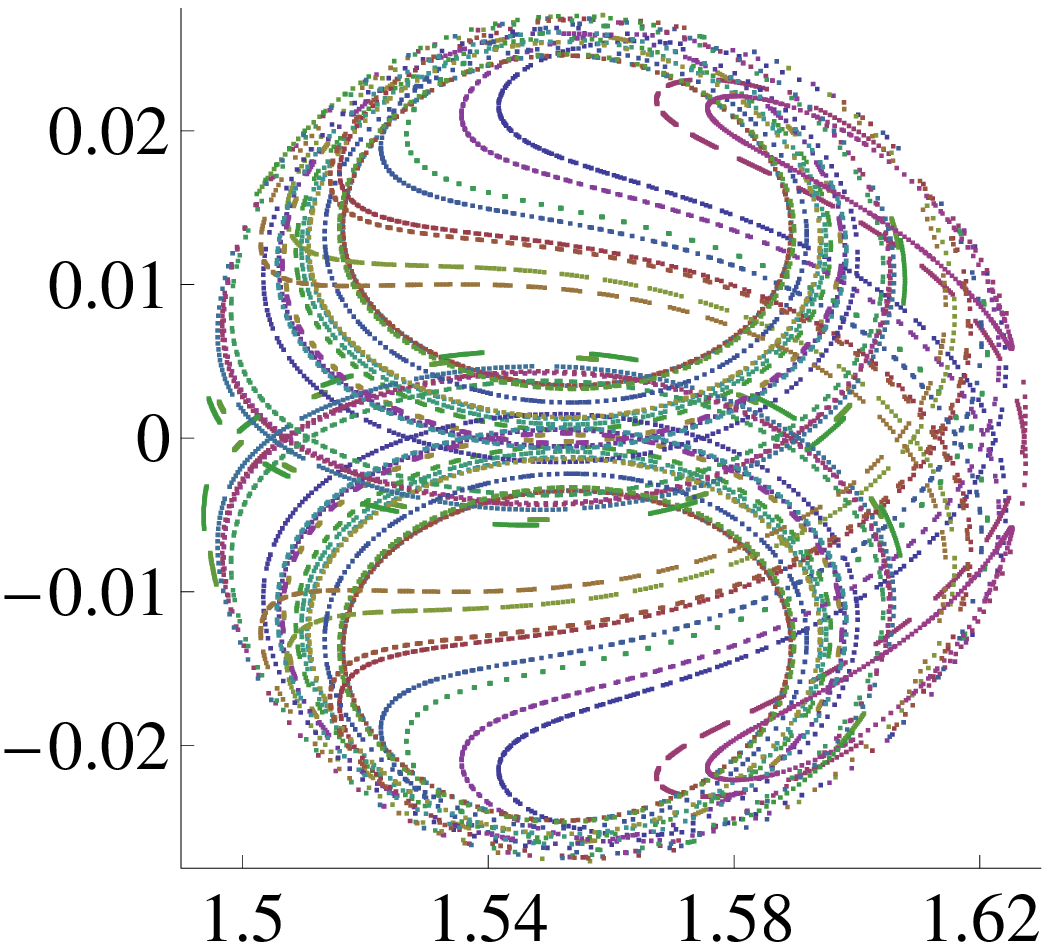}
 \end{center}
 \label{*****}
\end{minipage}
\hspace{5mm}
\begin{minipage}{0.3\hsize}
 \begin{center}
{(b) $E=0.947$}
 \includegraphics[width=5cm,clip]{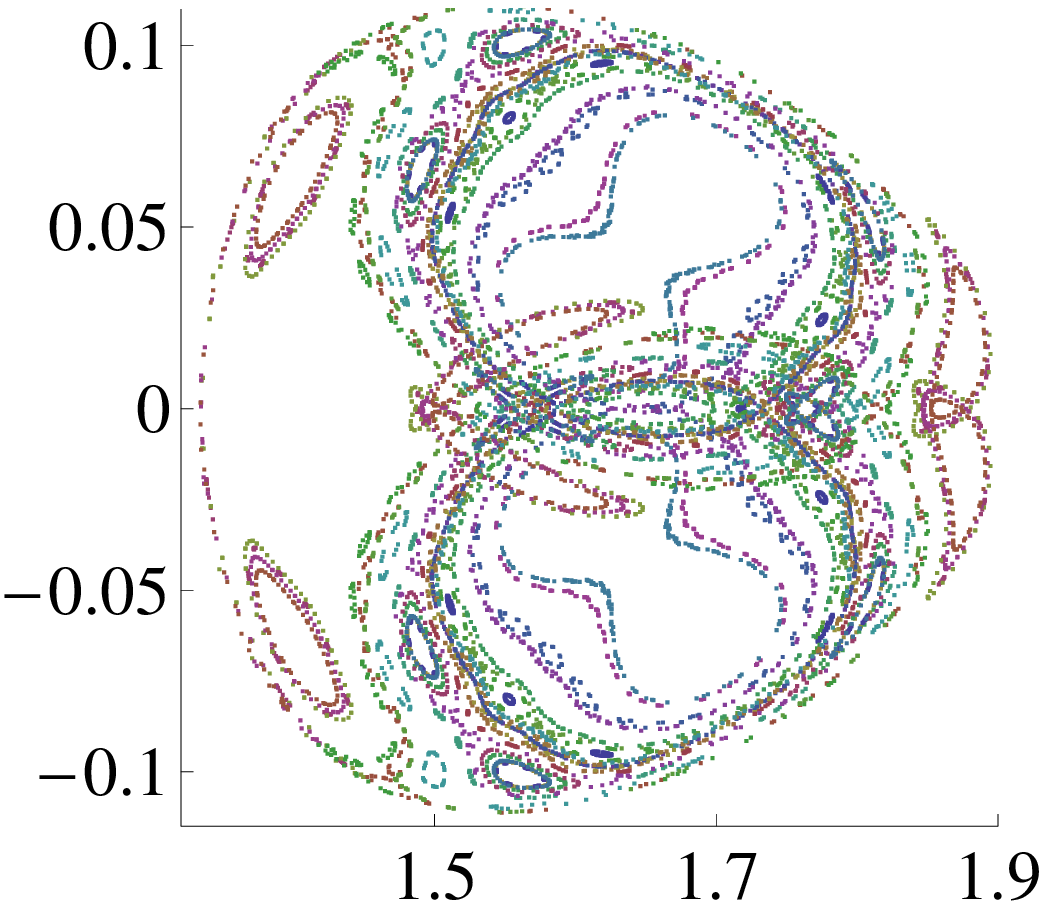}
 \end{center}
 \label{*****}
\end{minipage}
\hspace{5mm}
\begin{minipage}{0.3\hsize}
 \begin{center}
{(c) $E=0.952$}
 \includegraphics[width=5cm,clip]{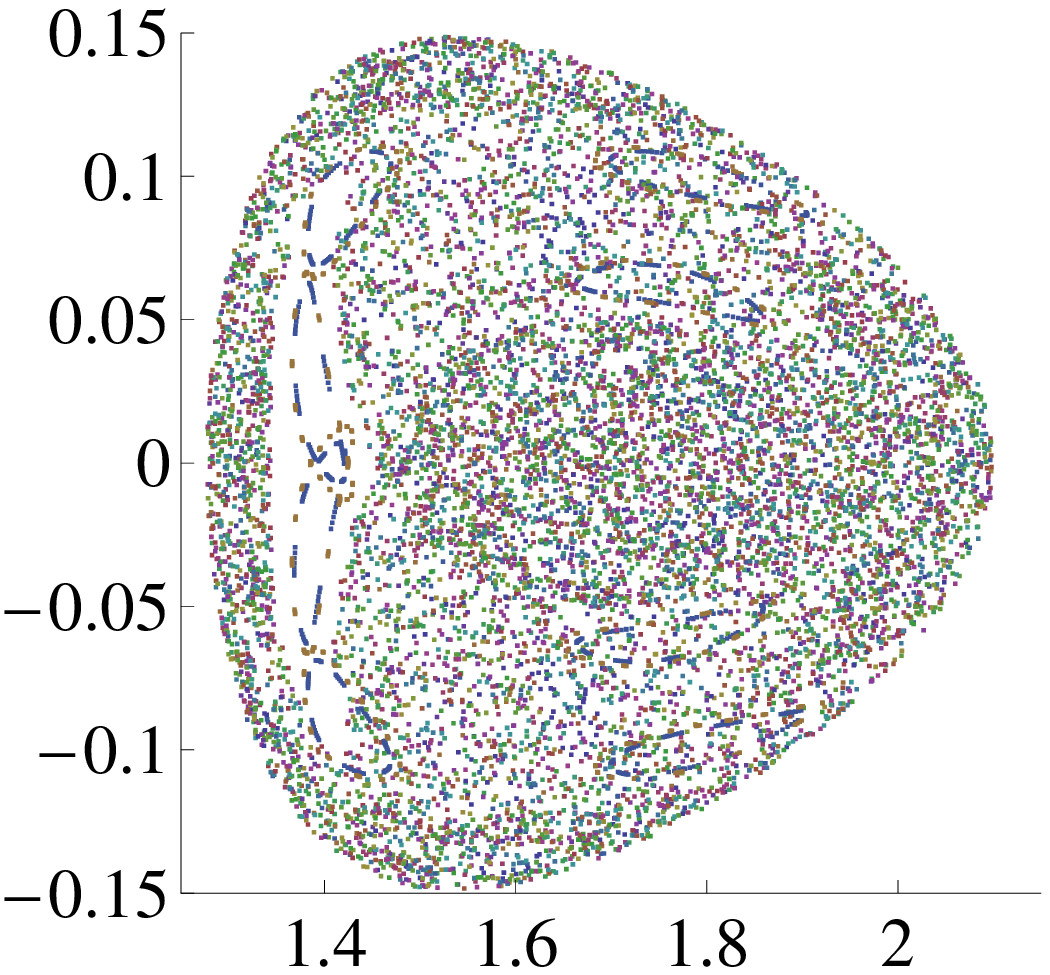}
 \end{center}
 \label{*****}
\end{minipage}
\end{tabular} 
 \caption{
\label{Poincare}
The Poincar\'e maps with the surface of section $\zeta=\zeta_{\rm s}=0.3$ are 
shown in the $\rho$-$p_{\rho}$ plane. 
The horizontal axis is $\rho$ and the vertical axis is $p_\rho$. 
The parameters are same as Fig.\ref{dynamical_orbits}. 
Thirty orbits with different initial conditions are superposed in 
each panel.  
}
\end{figure}

To inspect the trajectories in the phase space $(\zeta, \rho, p_\zeta, p_\rho)$, 
we use the Poincar\'e map. 
We plot intersections of a trajectory by the surface of section 
$\zeta=\zeta_{\rm s}$ 
with $p_\zeta>0$ on the two-dimensional $\rho$-$p_\rho$ plane 
(see Fig. \ref{Poincare}). 
In the low energy case, we see the plotted points 
lie on a closed curve in the $\rho$-$p_\rho$ plane. 
As the energy $E$ increases, the closed curve in the Poincar\'e map is 
modulated and broken. In the high energy case, 
%where the orbit can approach the saddle point of $U_{\rm eff}$, 
sections of a single trajectory fill a finite region. 
The behavior of the Poincar\'e map which depends on the energy of the particle 
is the same as the H\'enon-Heiles system\cite{Henon-Heiles}. 
The scattered points of the Poincar\'e map implies the particle motion is chaotic. 
Therefore, we can conclude that there is no additional constant of motion
except the energy and the angular momenta 
which are related to the isometries of the metric. 

%%%%%%%%%%%%%%%%%%%%%%%%%%%%%%%%%%%%%%%%%%%%%%%%%%%%%%%%%%%%%%%%%%%%%%%%%%%%%%%%%%
%\subsection*{Acknowledgements}
%%%%%%%%%%%%%%%%%%%%%%%%%%%%%%%%%%%%%%%%%%%%%%%%%%%%%%%%%%%%%%%%%%%%%%%%%%%%%%%%%%
This work is supported by the Grant-in-Aid for Scientific Research No.19540305.

%\newpage

%%%%%%%%%%%%%%%%%%%%%%%%%%%%%%%%%%%%%%%%%%%%%%%%%%%%%%%%%%%%%%%%%%%%%%%%%%%%%%%%%%

\end{document}